\documentstyle[12pt,epsf]{article}
\newcommand{\beq}{\begin{eqnarray}}
\newcommand{\eeq}{\end{eqnarray}}
\renewcommand{\Re}[1]{\textrm{Re}\!\left(#1\right)}

\newcommand{\abs}[1]{\left| #1 \right|}
\newcommand{\order}[1]{{\cal O}\!\left(#1\right)}
\renewcommand{\order}[1]{O\!\left(#1\right)}

\newcommand{\deltaGS}{\delta_{{\rm GS}}}
\begin{document}
\begin{titlepage}
\begin{flushright}
 KUNS-1880\\
 NIIG-DP-03-6
\end{flushright}

\begin{center}
\vspace*{5mm}
  
{\LARGE\bf
Non-perturbative K\"ahler Potential, Dilaton\\[7pt]
Stabilization and Fayet-Iliopoulos Term}
\vspace{12mm}

{\large
Tetsutaro~Higaki\rlap,\,\footnote{
E-mail address: tetsu@gauge.scphys.kyoto-u.ac.jp}
Yoshiharu~Kawamura\rlap,\,\footnote{
E-mail address: haru@azusa.shinshu-u.ac.jp}
Tatsuo~Kobayashi\,\footnote{
E-mail address: kobayash@gauge.scphys.kyoto-u.ac.jp}\\[5pt]
and~~Hiroaki~Nakano\,\footnote{
E-mail address: nakano@muse.sc.niigata-u.ac.jp}
}
\vspace{6mm}

{\it $^{\ast,\ddag}$Department of Physics, Kyoto University,
Kyoto 606-8502, Japan}\\[1mm]
{\it $^\dag$Department of Physics, Shinshu University,
Matsumoto 390-8621, Japan}\\[1mm]
{\it $^\S$Department of Physics, Niigata University,
Niigata 950-2181, Japan}

\vspace*{15mm}

\begin{abstract}
We study the dilaton stabilization
in models with anomalous $U(1)$ symmetry
by adding specific string-motivated, non-perturbative corrections
to the tree-level dilaton K\"{a}hler potential.
We find that the non-perturbative effects can stabilize the dilaton 
at a desirably large value. 
We also observe that the size of Fayet-Iliopoulos term
is reduced at the stabilized point.
\end{abstract}

\end{center}
\end{titlepage}


Dilaton and moduli fields play an important role in 
superstring theory as well as extra dimensional models.
Within the framework of 4D string models, couplings like 
gauge and Yukawa couplings are determined by 
vacuum expectation values (VEVs) of these fields.
In heterotic models, for example, the gauge coupling $g$ is 
determined as $1/g^2= \langle \Re{S} \rangle$ 
by the VEV of the dilaton field $S$.
However, in 4D models with $N=1$ supersymmetry (SUSY) 
these fields have perturbatively flat potential, 
and their VEVs are undetermined.
Thus, how to stabilize their VEVs is an important problem.
The non-perturbative superpotential due to gaugino 
condensations is a plausible origin for stabilizing their VEVs.
However, in the case with a single gaugino condensation 
and the tree-level K\"ahler potential, 
\begin{equation}
K_0(S+\bar S)={}-\ln(S + \bar S) \ ,
\label{K:tree}
\end{equation}
the dilaton VEV can not be stabilized 
at a finite value, but runs away to infinity.

Several models have been proposed to stabilize the dilaton VEV.
The models with double or more gaugino condensations, 
i.e. the so-called racetrack models, can 
stabilize the dilaton VEV \cite{Krasnikov:jj}.
The problem of the racetrack type models is that
the stabilized value of the dilaton tends to be too small
compared with the value $\Re{S} = 1/g^2 \approx 2$, 
which is suggested by the unified gauge coupling 
in the minimal supersymmetric standard model.
A certain degree of fine-tuning is necessary
to realize the dilaton stabilization at weak coupling region.

Another possibility for the dilaton stabilization is 
to assume non-perturbative K\"ahler potential 
of the dilaton field \cite{Shenker:1990uf,Banks:1994sg},
as was studied in Refs.~\cite{Banks:1994sg}-\cite{Choi:1998nx}.
With a certain form of non-perturbative K\"ahler potential,
a single gaugino condensation can stabilize the dilaton 
at a finite value.
Moreover, the dilaton VEV of $\order{1}$ can be realized 
for a reasonable choice of parameters,
although one has still to fine-tune parameters 
so that the tree-level vacuum energy vanishes.

On the other hand, it is usually true that
$D$-terms in the scalar potential do not
play any essential role on dilaton stabilization,
because the dilaton field appears as an overall factor in $D$-terms.
There can happen, however, an exception, that is, the case with
$D$-term for an anomalous $U(1)$ symmetry.
Most of 4D string models have anomalous $U(1)$ 
symmetries \cite{DSW,Kobayashi:1996pb,typeI},
whose anomalies can be cancelled by the Green-Schwarz (GS) mechanism.
In heterotic models,  
the dilaton field transforms nonlinearly 
like $S \rightarrow S+2i\deltaGS\Lambda_A$ 
under anomalous $U(1)$ transformation
$V_{A} \rightarrow V_{A} +i\Lambda_A -i\bar{\Lambda}_A$,
where $\deltaGS$ is a GS coefficient and $V_A$ is 
the anomalous $U(1)$ vector multiplet. 
It follows that the dilaton K\"ahler potential 
is a function $K(s)$ of gauge-invariant combination
$s \equiv S + \bar{S} - 2\deltaGS{}V_{A}$.
Accordingly, the anomalous $U(1)$ $D$-term contains
the Fayet-Iliopoulos (FI) term 
\begin{equation}
\xi = \deltaGS \langle K_S \rangle M^2 \ ,
\label{FI-term}
\end{equation}
where $M$ is the reduced Planck scale
and $K_S$ is the first derivative of the dilaton K\"ahler potential.
If we take the tree-level K\"ahler potential $K_0(s)$
and assume that $\Re{S}=\order{1}$, 
we have $\xi^{1/2}/M = 10^{-1}\,\hbox{--}\,10^{-2}$. 
(Hereafter we take the $M=1$ unit.)
In general, the magnitude of the FI term depends on the dilaton VEV
as well as the form of dilaton K\"ahler potential.
Therefore, the anomalous $U(1)$ $D$-term can play a nontrivial role
in dilaton stabilization,
as was suggested before in Refs.~\cite{Banks:1995ii,Dine:1998qr}.

The dilaton-dependent FI term has also several phenomenologically 
interesting aspects.
{}For example, the ratio of the FI term to the Planck mass 
squared can be an origin of coupling hierarchies \cite{FN,IR}.
The FI term can also be used 
to break SUSY \cite{SUSY-breaking:mech}--\cite{SUSY-breaking2}
as well as to mediate SUSY-breaking effects
to scalar mass terms \cite{DA}--\cite{Higaki:2003ig}.
{}Furthermore, in the $D$-term inflation scenario, 
the FI term is a dominant term in the vacuum energy driving 
the inflation \cite{inflation}.\footnote{
See Refs~\cite{Halyo:1999bq,Kobayashi:2003rx} for 
$D$-term inflation scenarios in type I models.
}
In these applications, the size of the FI term, 
which is determined as eq.~(\ref{FI-term}) in the heterotic case,
is quite important.

In this paper, we study the dilaton stabilization mechanism 
in which a dominant role is played 
by the dilaton-dependent FI term (\ref{FI-term})
due to non-perturbative K\"ahler potential.
In this scenario, 
the dilaton VEV can easily be stabilized 
at weak coupling, $\Re{S}=\order{1}$, as we will see below.
Similar studies have been done 
in Refs.~\cite{Arkani-Hamed:1998nu,Barreiro:1998nd}, 
where the superpotential due to gaugino condensation is 
also added to stabilize the dilaton VEV.
In our case, however, 
we do not assume such dilaton-dependent superpotential.
This means that the dominant part of scalar potential 
$V$ is given by $V \sim (\deltaGS{} K_S)^2$.
As a result, the dilaton VEV is stabilized 
around the point satisfying $K_S = 0$.
This minimum corresponds to the point discussed before 
from the viewpoint of maximally enhanced symmetry \cite{Dine:1998qr}.
Moreover, 
we will present an example of dilaton-dependent superpotential
that does not spoil the dilaton stabilization
through the anomalous $U(1)$ $D$-term
so that the resulting FI term has a suppressed value
compared with the value expected from the tree-level K\"ahler potential.

Basically it is difficult to stabilize the dilaton
only through the $D$-term scalar potential
if the K\"ahler potential takes the tree-level form (\ref{K:tree}).
To realize it,
we assume that non-perturbative effects generate another 
term in the dilaton K\"ahler potential.
Of course, it is not clear, at present, 
which type of terms would be generated by non-perturbative physics.
Therefore, for illustrating purpose, we use the following 
Ansatz for non-perturbative potential \cite{Casas:1996zi},
\beq
K_{{\rm np}}(S+\bar S )
 = d\left(S + \bar S\right)^{p/2}e^{-b (S+\bar S)^{1/2}} \ ,
\eeq
where $d,p$ and $b$ are real constants.
It is required that $b >0$, for
the non-perturbative term must vanish in the weak coupling limit, 
$\Re{S} = 1/g^{2} \rightarrow \infty$.
Then, in models with an anomalous $U(1)_A$,  
we consider the total K\"ahler potential of dilaton,
\beq
K^{{\rm (I)}}(s) =  K_{0}(s) + K_{{\rm np}}(s) \ .
\eeq
Alternatively, the total dilaton K\"ahler potential of the form
\beq
K^{{\rm (I\!I)}}(s)
= \ln \left(e^{K_{0}(s)}+e^{K_{{\rm np}}(s)} \right)
\eeq
has also been discussed in the literature.
We also give comments on the case with $K^{{\rm (I\!I)}}(s)$.

Now let us explain our setting.
The total K\"{a}hler potential takes the form
\beq
K=K\!\left(S+\bar{S}-2\deltaGS{}V_A\right)
 +K\!\left(\Phi^i,\bar{\Phi}^{\bar{i}}\right)
 +\sum_{\kappa}K_{\kappa{\bar{\kappa}}}
  \!\left(\Phi^i,\bar{\Phi}^{\bar{i}}\right)
  \bar{\phi}^{\bar{\kappa}}e^{2q^{A}_{\kappa}V_{A}}\phi^{\kappa}
 + \cdots \ ,
\eeq
where the first term is the dilaton K\"ahler potential 
$K^{{\rm (I)}}$ or $K^{{\rm (I\!I)}}$.
In the second and third terms,
$\Phi^i$ are gauge singlet moduli fields other than the dilaton field,
and $\phi^{\kappa}$ stand for matter fields 
with $U(1)_A$ charge $q^{A}_{\kappa}$.
The ellipsis denotes terms including gauge multiplets 
other than $U(1)_A$ and higher order terms of $\phi^{\kappa}$.
{}For superpotential $W$,
we first consider the model in which
$W$ does not include the dilaton field,
\beq
W=W\!\left(\Phi^i,\phi^{\kappa}\right) \ ,
\eeq
unlike the non-perturbative term generated by gaugino condensation.
This is an important assumption 
and we will come back to this point later.

Under the above setting,
the scalar potential is given by 
\beq
V&=& e^{K}
 \left[\frac{1}{K_{S\bar{S}}}\abs{K_{S}W}^2 
 +(K^{-1})^{I\bar{J}}
  \Bigl(K_{I}W+W_{I}\Bigr)
  \Bigl(K_{\bar{J}}\bar{W}+\bar{W}_{\bar{J}}\Bigr)
 -3\abs{W}^2 \right] 
\nonumber\\
 &&{}+\frac{1}{2\Re{S}}
      \left(\deltaGS{}K_{S}-\sum_{\kappa}q^{A}_{\kappa}
      K_{\kappa{\bar{\kappa}}}\abs{\phi^{\kappa}}^2\,\right)^2
 + \cdots \ ,
\eeq
where 
$K_{S \bar S}$ is the K\"ahler metric of the dilaton field,
and subindices $I,J$ represent derivatives 
with respect to the $\Phi^{i}$ or $\phi^{\kappa}$.
Here the ellipsis denotes $D$-terms other than the $U(1)_A$ $D$-term.
A solution of the stationary condition 
$\partial V /\partial S =0$ is given by 
\beq
K_S =0 \ , \qquad 
\Delta\equiv\sum_{\kappa}q^{A}_{\kappa}
K_{\kappa{\bar{\kappa}}}\abs{\phi^{\kappa}}^2=0 \ .
\label{conditionoffield}
\eeq
The first equation is the condition of vanishing FI term,
from which the dilaton is stabilized as we shall see shortly.
We have assumed that 
the second condition in eq.~(\ref{conditionoffield})
also satisfies  $F$-flatness conditions.
Actually, this solution corresponds to vanishing $F$-term of 
$S$ and vanishing $U(1)_A$ $D$-term, 
so that SUSY is unbroken in the dilaton sector.
At this point (\ref{conditionoffield}), 
the second derivative of $V$ is written as 
\begin{equation}
\left.
\frac{\partial^2 V}{\partial S \partial \bar S} 
\right\vert_{K_S = \Delta =0}
 = \left\langle K_{S \bar S} V + 2K_{S \bar S} e^K\abs{W}^2
  +\frac{\deltaGS^2 K_{S \bar S}^2}{\Re{S}} \right\rangle \ .
\label{dV-2}
\end{equation}
On the right hand side of this equation,
the first term can be neglected when 
the (tree-level) vacuum energy is taken to be approximately zero.
(Note that the vacuum energy contribution from the dilaton sector 
vanishes at $K_S =\Delta=0$.)
Moreover, the second derivative $K_{S\bar{S}}$ must be positive
because it determines a normalization of kinetic term of the dilaton.
We find that the right hand side of eq.~(\ref{dV-2}) are positive
at $K_S =\Delta=0$, and thus the equation~(\ref{conditionoffield}) 
corresponds to a local minimum of the scalar potential $V$.

Let us discuss a concrete example.
We consider the K\"ahler potential $K^{{\rm (I)}}$.
Its first derivative with respect to the dilaton is 
obtained as 
\beq
K^{{\rm (I)}}_{S}(s)
 ={}-\frac{1}{s}
    +\frac{d}{2}\,s^{p/2-1}e^{-bs^{1/2}}
  \left[p-bs^{1/2}\right] \ .
\label{KI'}
\eeq
The solutions to the equation $K^{{\rm (I)}}_{S}=0$
behave differently for $d<0$ case and $d>0$ case.
When $p$ and $b$ are positive and fixed,
the $d < 0$ case can lead to larger value of 
$\Re{S}$ than the $d>0$ case.
{}For example, in the case with $p=b=1$ and $d =\!{}-e^2$, 
the dilaton VEV is stabilized as $\Re{S}=2$, while 
we obtain $\Re{S} = 0.125$ in the case with 
$p=b=1$ and $d =8 e^{1/2}$.
Since we are interested in the solution $\Re{S}=\order{1}$,
we will mainly consider the case with $d<0$
and give a brief comment for $d>0$ later.

\begin{figure}
\epsfxsize=0.7\textwidth
\centerline{\epsfbox{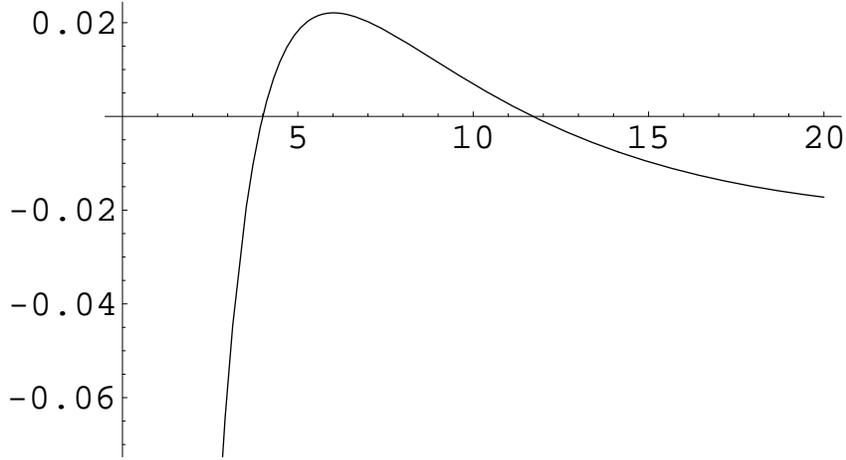}}
\caption{
$K^{{\rm (I)}}_{S}$ as a function of $s=2\Re{S}$.
The parameters are $p=b=1$ and $d=\!{}-e^2$.
}
\label{figure1}
\end{figure}

\begin{figure}
\epsfxsize=0.66\textwidth
\centerline{\hspace*{18pt}
\epsfbox{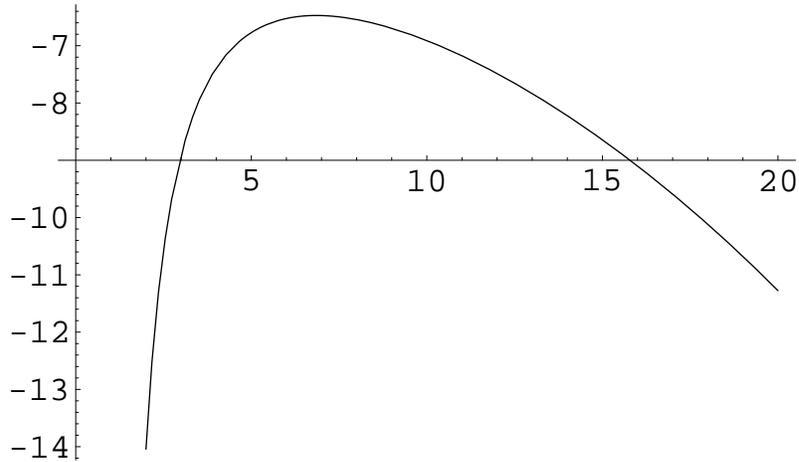}}
\caption{
The curve of $d$ (the vertical axis)
against  $s=2\Re{S}$ (the horizontal axis)
which satisfy $K^{{\rm (I)}}_{S}=0$ for $p=b=1$. 
{}For $s > 7.8$, we have $K_{S \bar S} <0$ and 
such part of this curve does not correspond to a physical solution.
}
\label{figure2}
\end{figure}

{}Figure~\ref{figure1} shows $K^{{\rm (I)}}_S$ 
for $p=b=1$ and $d=\!{}-e^2$.
We see that there are two solutions to $K^{{\rm (I)}}_S =0$
(except the runaway one);
one corresponds to the solution with $K^{{\rm (I)}}_{S \bar S} >0$ 
while the other gives $K^{{\rm (I)}}_{S \bar S} <0$.
Thus the physical solution is given by 
$\Re{S}=2$ as mentioned above.
We also show in Figure~\ref{figure2} how
the stabilized dilaton VEV depends on the parameter $d<0$.
As $|d|$ becomes large, the stabilized value becomes small.
In the limit $|d| \rightarrow \infty$, 
the stabilized value $\Re{S}$ comes close to $1/2$.  
On the other hand, as $|d|$ becomes small, 
the stabilized value $\Re{S}$ becomes large.
However, for $d >\!{}-6.5$, we have no solution to $K^{{\rm (I)}}_S =0$.
The maximum value of the dilaton VEV 
is $\Re{S}\approx 3.4$ for $d\approx\!{}-6.5$.
We note that 
in general the second derivative $K^{{\rm (I)}}_{S \bar S}$ 
is suppressed slightly.
{}For example, we have $K_{S\bar{S}}=1/32$ for $d=\!{}-e^2$.

{}For other values of $p$ and $b$, we obtain qualitatively 
the same results.
The limit $|d| \rightarrow \infty $ corresponds to 
the minimum of $\Re{S}$, which is obtained as $\Re{S}=p^2/(2b^2)$.
As $d$ decreases, the stabilized value increases.

Here we give a comment on the case with $d>0$.
{}For $p$ and $b$ fixed positively, 
as $d$ decreases, the stabilized value of $\Re{S}$ 
increases, but it can not be larger than $p^2/(8b^2)$.
Thus, for $d>0$ we have $\Re{S}=\order{1}$ 
for a large ratio of $p^2/(8b^2)$ and a small value of $d$.

Similarly we can discuss the dilaton stabilization for 
$K^{{\rm (I\!I)}}$.
Its first derivative $K^{{\rm (I\!I)}}_{S}$ is calculated to be
\beq
K^{{\rm (I\!I)}}_{S}=
\frac{1}{1+s\exp\left(ds^{p/2}e^{-b s^{1/2}}\right)}
\left[-\frac{1}{s} +\frac{d}{2}\,s^{p/2}e^{-bs^{1/2}}
\left(p-bs^{1/2}\right)
\exp\left(ds^{p/2}e^{-bs^{1/2}}\right)
\right] \ .
\label{KII'}
\eeq
{}For example when $p=b=\!{}-d=1$, 
the equation $K^{{\rm (I\!I)}}_{S}=0$ 
is satisfied by $\Re{S}=3.9$, where we have $K_{S\bar{S}}=0.13$.

So far, we have considered the model 
without dilaton-dependent superpotential.
In that case, the minimum of the scalar potential is determined 
by $K_{S}=0$ corresponding to vanishing FI term.
On the other hand, if a dilaton-dependent term
is generated non-perturbatively in the superpotential,
one may expect that such term would drastically change the situation,
that is, the dilaton VEV would no longer be determined 
by the anomalous $U(1)$ $D$-term.
This is not necessarily the case, however.
We now present a class of models in which
the superpotential contains a dilaton-dependent term,
but the dilaton VEV is dominantly determined 
by the anomalous $U(1)$ $D$-term.
In fact, a sub-dominant effect from the superpotential
slightly shifts the minimum from the point $K_{S}=0$,
as we shall see shortly.

Here we consider a toy model with $SU(2)\times U(1)_{A}$ gauge group.
The model has four $SU(2)$ doublet chiral superfields 
$Q_{i}^{a}$ ($i=1,\cdots,4$; $a=1,2$) 
which have anomalous $U(1)_{A}$ charges $q_i$
with $\sum_{i}q_i \neq 0$.
In this case, the $SU(2)$ strong dynamics deforms 
the moduli space of vacua into \cite{Seiberg:1994bz}
\beq
{\mathrm Pf}\left(M_{ij}\right) = \exp\!\left(\!{}-8\pi^{2}S\right)~ ,
\label{QMS}
\eeq
where $M_{ij}$ is the meson operator corresponding to $Q_iQ_j$.
The right hand side corresponds to $\Lambda^4$, where 
$\Lambda=\exp[-2\pi^2 S]$ is the dynamical scale (in the $M=1$ unit).
Suppose that the superpotential includes only the term with 
a Lagrange multiplier that enforces the above constraint (\ref{QMS}).
{}Furthermore, 
we assume K\"{a}hler potentials of $M_{ij}$ to be 
$K(M_{ij},\bar M_{ij} )= (M_{ij} \bar M_{ij})^{1/2}$ 
for simplicity.
Then, the anomalous $U(1)_{A}$ $D$-term takes the form
\beq
D =  \deltaGS{}K_S - \sum
\frac{q_{ij}}{2}\left(M_{ij} \bar M_{ij}\right)^{1/2}~,
\eeq
where $q_{ij} = q_i + q_j$.

Now, we may estimate the minimum of the scalar potential by solving
\beq
\deltaGS{}K_S
 = \sum \frac{q_{ij}}{2}\left(M_{ij}\bar M_{ij}\right)^{1/2} \ .
\label{cond-vac}
\eeq
Combining eq.~(\ref{cond-vac}) with the quantum constraint (\ref{QMS}),
we obtain
\begin{equation}
K_S = \order{\exp\!\left[\beta-4\pi^2 \Re{S}\right]} \ ,
\label{SFI}
\end{equation}
where we have defined $\deltaGS \equiv e^{-\beta}$
and assumed that $q_{ij}=\order{1}$.
Normally we have $\beta=\order{1}$ since
$\deltaGS= 10^{-1}\,\hbox{--}\,10^{-2}$ in the unit $M=1$.
If the stabilized value before adding the superpotential
is given by $\Re{S}=\order{1}$,
the right hand side in eq.~(\ref{SFI}) is sufficiently suppressed
as long as $\beta=\order{1}$.
If this is the case,
we may consistently approximate the minimum condition
by $K_S \approx 0$ as before.
This situation does not change even for $\beta=\order{10}$ 
because $4\pi^2\Re{S} \gg \beta$.

It is important, however, to notice that
the FI term $\xi$ does not vanish exactly.
In the above toy model, it is estimated as 
\beq
|\xi| = \abs{\deltaGS{}K_S} M^2 
 =\order{ {M^2}\exp\!\left(\!{}-8\pi^2 \right)}
\sim \order{10^2}\,\mathrm{GeV}^2 \ .
\eeq
when $\Re{S}=2$.
Thus the FI term is nonvanishing, 
but quite suppressed in this model.
If we consider a  model with larger rank of gauge group, 
the dynamical scale $\Lambda$ can be larger. Accordingly 
a larger FI term $\xi=\order{\Lambda^2}$ can be generated.
{}For example, in the model which has $SU(7)$ gauge group with seven
flavors and $\Re{S}=2$, we obtain 
the dynamical scale $|\Lambda| \approx 10^{13}$ GeV.
In general, this type of models lead, up to $U(1)$ charges, to 
\begin{equation}
\frac{|\xi|}{M^2} = \abs{\deltaGS{}K_S}
= \exp\!\left[{}-\frac{8\pi^2}{b'}\,2\Re{S}\right] \ ,
\label{s-model2}
\end{equation}
where $b'$ is the one-loop gauge beta-function coefficient 
in the model with quantum moduli space.
We also note that the stabilized VEV of $2\Re{S}$ is slightly shifted 
from the value $s_0$ of previous case satisfying $K_S(s_0)=0$ exactly.
Such shift $\delta{}s$ is negligible as long as
\begin{equation}
\frac{8 \pi^2}{b'\deltaGS{}K_{S \bar S}(s_0)}\,
\exp\!\left(\!{}-\frac{8\pi^2}{b'}\,s_0 \right) \ll{} 1 \ .
\end{equation}
Otherwise, the shift is not small, and we have to fully solve 
the stationary condition of the scalar potential.


To summarize, we have studied the dilaton stabilization 
in the model with the non-perturbative
dilaton K\"{a}hler potential and anomalous $U(1)$ gauge symmetry.
It is found that non-perturbative effects can stabilize the dilaton 
at a finite value of $\order{1}$.
Another interesting property of this stabilization mechanism 
is that one can reduce the order of magnitude of FI term. 
We give a toy model in which
small dynamical scale and FI term are generated.
If gauge group is larger, they can become larger.
That would have interesting applications e.g. for 
the $D$-term inflation scenario.
{}Finally we add that in the models discussed here, 
SUSY is not broken in the dilaton sector, 
and the tree-level vacuum energy contribution from this sector vanishes.
In order to break SUSY, we must take into account effects 
from other moduli fields or tree-level superpotential.

\section*{Acknowledgment}

One of the authors (T.~K.) would like to thank Kiwoon Choi 
for useful discussions.
The authors would like to thank the organizers of 
Summer Institute 2003, Fuji-Yoshida, Yamanashi, Japan, where 
a part of this work was studied.
Y.~K.\/ is supported in part by the Grant-in-Aid 
for Scientific Research from Ministry of Education, Science, Sports
and Culture of Japan (\#13135217) and (\#15340078).
T.~K.\/ is supported in part by the Grant-in-Aid for 
Scientific Research from Ministry of Education, Science, 
Sports and Culture of Japan (\#14540256). 
T.~H.\/ and T.~K.\/ are supported in part by the Grant-in-Aid for 
the 21st Century COE ``The Center for Diversity and 
Universality in Physics'' from Ministry of Education, Science, 
Sports and Culture of Japan.


\begin{thebibliography}{99}

\bibitem{Krasnikov:jj}
N.~V.~Krasnikov,
Phys.\ Lett.\ B {\bf 193}, 37 (1987); \\
L.~J.~Dixon,
SLAC-PUB-5229
{\it Invited talk given at 15th APS Div. of Particles and Fields
  General Mtg., Houston,TX, Jan 3-6, 1990};\\
T.~R.~Taylor,
Phys.\ Lett.\ B {\bf 252}, 59 (1990);\\
J.~A.~Casas, Z.~Lalak, C.~Munoz and G.~G.~Ross,
Nucl.\ Phys.\ B {\bf 347}, 243 (1990);\\
B.~de Carlos, J.~A.~Casas and C.~Munoz,
Nucl.\ Phys.\ B {\bf 399}, 623 (1993);\\
M.~Dine and Y.~Shirman,
Phys.\ Rev.\ D {\bf 63}, 046005 (2001).


\bibitem{Shenker:1990uf}
S.~H.~Shenker,
RU-90-47
{\it Presented at the Cargese Workshop on Random Surfaces, 
Quantum Gravity and Strings, Cargese, France, May 28 - Jun 1, 1990}



\bibitem{Banks:1994sg}
T.~Banks and M.~Dine,
Phys.\ Rev.\ D {\bf 50}, 7454 (1994).


\bibitem{Binetruy:1996xj}
P.~Binetruy, M.~K.~Gaillard and Y.~Y.~Wu,
Nucl.\ Phys.\ B {\bf 481}, 109 (1996).


\bibitem{Casas:1996zi}
J.~A.~Casas,
Phys.\ Lett.\ B {\bf 384}, 103 (1996).

\bibitem{Barreiro:1997rp}
T.~Barreiro, B.~de Carlos and E.~J.~Copeland,
Phys.\ Rev.\ D {\bf 57}, 7354 (1998).

\bibitem{Choi:1998nx}
K.~Choi, H.~B.~Kim and H.~D.~Kim,
Mod.\ Phys.\ Lett.\ A {\bf 14}, 125 (1999).


\bibitem{DSW}
E.~Witten, Phys.~Lett. {\bf 149B}, 351 (1984);\\
M.~Dine, N.~Seiberg and E.~Witten, 
Nucl.~Phys. {\bf B289}, 589 (1987);\\
W.~Lerche, B.E.W.~Nilsson and A.N.~Schellekens,
Nucl.~Phys. {\bf B289}, 609 (1987).


\bibitem{Kobayashi:1996pb}
T.~Kobayashi and H.~Nakano,
Nucl.\ Phys.\ B {\bf 496}, 103 (1997)
;\\
G.B.~Cleaver and A.E.~Faraggi, 
Int.~J.~Mod.~Phys. {\bf A14}, 2335 (1999).



\bibitem{typeI}
L.E.~Ib\'a\~nez, R.~Rabad\'an and A.M.~Uranga,
Nucl.~Phys. {\bf B542}, 112 (1999);\\
Z.~Lalak, S.~Lavignac and H.P.~Nilles, 
Nucl.~Phys. {\bf B559}, 48 (1999).

\bibitem{Banks:1995ii}
T.~Banks and M.~Dine,
Phys.\ Rev.\ D {\bf 53}, 5790 (1996).

\bibitem{Dine:1998qr}
M.~Dine, Y.~Nir and Y.~Shadmi,
Phys.\ Lett.\ B {\bf 438}, 61 (1998) .

\bibitem{FN}
C.D.~Froggatt and H.B.~Nielsen, Nucl.~Phys. {\bf B147}, 277 (1979).

\bibitem{IR}
L.E.~Ib\'a\~nez and G.G.~Ross, Phys.~Lett. {\bf B332}, 100 (1994).

\bibitem{SUSY-breaking:mech}
G.~Dvali and A.~Pomarol, 
Phys.~Rev.~Lett. {\bf 77}, 3728 (1996);\\
P.~Binetruy and E.~Dudas, Phys.~Lett. {\bf B389}, 503 (1996).

\bibitem{Arkani-Hamed:1998nu}
N.~Arkani-Hamed, M.~Dine and S.~P.~Martin,
Phys.\ Lett.\ B {\bf 431}, 329 (1998).

\bibitem{Barreiro:1998nd}
T.~Barreiro, B.~de Carlos, J.~A.~Casas and J.~M.~Moreno,
Phys.\ Lett.\ B {\bf 445}, 82 (1998).



\bibitem{SUSY-breaking2}
A.~Kageyama, H.~Nakano, T.~Ozeki and Y.~Watanabe,
Prog.~Theor.~Phys. {\bf 101}, 439 (1999).

\bibitem{DA}
H.~Nakano, arXiv:hep-th/9404033;  
Prog. Theor. Phys. Suppl. {\bf 123}, 387 (1996).

\bibitem{KK}
Y.~Kawamura and T.~Kobayashi, Phys. Lett. {\bf B375}, 141 (1996) 
[Erratum: {\bf B388}, 867 (1996)];
Phys. Rev. {\bf D56}, 3844 (1997).

\bibitem{DPS}
E.~Dudas, S.~Pokorski and C.A.~Savoy, 
Phys. Lett. {\bf B369}, 255 (1996);\\
E.~Dudas, C.~Grojean, S.~Pokorski and C.A.~Savoy, 
Nucl.~Phys. {\bf B481}, 85 (1996).

\bibitem{K}
Y.~Kawamura, Phys. Lett. {\bf B446}, 228 (1999).


\bibitem{Higaki:2003ig}
T.~Higaki, Y.~Kawamura, T.~Kobayashi and H.~Nakano,
arXiv:hep-ph/0308110.


\bibitem{inflation}
P.~Binetruy and G.~Dvali,
Phys.~Lett. {\bf B388} (1996) 241;\\
E.~Halyo, Phys.~Lett. {\bf B387} (1996) 43.


\bibitem{Halyo:1999bq}
E.~Halyo,
Phys.\ Lett.\ B {\bf 454}, 223 (1999).


\bibitem{Kobayashi:2003rx}
T.~Kobayashi and O.~Seto,
arXiv:hep-ph/0307332, to be published in Phys. Rev. D.


\bibitem{Seiberg:1994bz}
N.~Seiberg,
Phys.\ Rev.\ D {\bf 49}, 6857 (1994);\\
K.~A.~Intriligator and N.~Seiberg,
Nucl.\ Phys.\ Proc.\ Suppl.\  {\bf 45BC}, 1 (1996) .







\end{thebibliography}
\end{document}